\documentclass[10pt]{article}
\usepackage[dvips]{graphicx}

\usepackage{epsfig}
\usepackage{graphicx}
\usepackage{indentfirst}
\usepackage{amsmath}
\usepackage{amsfonts}
\usepackage{amssymb}
\usepackage{hyperref}
 \usepackage{latexsym}
\usepackage{color}

\begin{document}
\begin{center}
{\Large\bf Locally Rotationally Symmetric Bianchi Type-I cosmological model in $f(T)$ gravity: from early to Dark Energy dominated universe}\\

{\small  M. E. Rodrigues $^{(a,b)}$}\footnote{E-mail
address: esialg@gmail.com},
{\small    I. G. Salako $^{(c)}$}\footnote{E-mail address:
ines.salako@imsp-uac.org}, 
{\small    M. J. S. Houndjo $^{(c)(d)}$}\footnote{E-mail address:
sthoundjo@yahoo.fr} and
{\small  J. Tossa $^{(c)}$}\footnote{E-mail address: joel.tossa@imsp-uac.org},
 \\ 
$^{a}$\,{\it\ Faculdade de F\'{\i}sica, Universidade Federal do Par\'{a}, 66075-110, Bel\'em, Par\'{a}, Brazil}\\
$^{b}$\,{\it\ Faculdade de Ci\^{e}ncias Exatas e Tecnologia, Universidade Federal do Par\'{a} - Campus Universit\'ario de Abaetetuba, CEP 68440-000, Abaetetuba, Par\'{a}, Brazil}\\

$^{c}${\it Institut de Math\'{e}matiques et de Sciences Physiques (IMSP) - 01 BP 613 Porto-Novo, B\'{e}nin}\\
$^{d}${\it Facult\'e des Sciences et Techniques de Natitingou - Universit\'e de Parakou - B\'enin     }\\

\end{center}
\begin{abstract}
We study the locally rotational symmetry Bianchi type-I dark energy model in the framework of $f(T)$ theory of gravity, where $T$ denotes the torsion scalar. A viable cosmological model is undertaken  and the isotropization of this latter is checked, yielding a result that reflects the real evolution of our universe. Moreover, still in the anisotropic optic, a more complicated $f(T)$ model is obtained from the cosmological reconstruction scheme and the analysis shows that the universe is more anisotropic at the beginning if the terms of higher order in $T$ are not considered. This means that the non-linear model should be favoured by observational data. 
\end{abstract}
Pacs numbers: 98.80.-k, 04.50.Kd, 04.20.Jb
\section{Introduction}
The discovery of the accelerated expansion of the late-time universe gives rise to a wide spectrum of various theories in order to explain this  acceleration. In the general gravity (GR) scheme, the dark energy, known to have an exotic property such as negative pressure, is considered to be a first candidate to this acceleration of the universe \cite{universe}.
Still in this spirit, and as alternatives to the GR, several theories based on the modification of the  the curvature scalar have been developed with interesting results. The most important ones are the $f(R)$ \cite{r1}-\cite{r4}, $f(G)$ \cite{mj1}-\cite{mj5},  $f(R,G)$ \cite{rg1}-\cite{rg8}, $f(R,\mathcal{T})$ \cite{ma1}-\cite{ma6}, where $R$, $G$ and $\mathcal{T}$ are the curvature scalar, the Gauss-Bonnet invariant and the trace of the energy-momentum tensor, respectively.\par
Nevertheless, there exists another alternative to the dark energy, being the modification of the teleparallel theory equivalent to the GR (TEGR), namely $f(T)$ theory of gravity, where $T$ denotes the torsion scalar. Instead of the Levi-Civita's connection, as used in the GR and its different modified versions, the connection under consideration in $f(T)$ is that of Weitzenbok.  This theory has been introduced first by Ferraro \cite{ferraro} where they explained the $ UV$ modifications to GR and also the inflation. Afterwards, Ferraro and Bengochea considered the same models in the context of late cosmology to describe dark energy \cite{ferraro2}. Other authors developed various cosmological and gravitational ideas, still in $f(T)$ theory of gravity. In \cite{boehmer}, Tamanini and Boehmer investigate the importance of choosing good tetrads for the study of the field equations of $f(T)$ gravity. It is important to recall that the use of specific tetrads is extremely crucial of having a reasonable feature in this theory. Note that when the set of tetrads is not suitably chosen, the theory falls into the usual teleparallel one and there will be no reason of why having modified the teleparallel theory. This important scheme of choosing a suitable set of tetrads is also undertaken in this manuscript and presented in the section $3$, where it will be observed that the algebraic function $f(T)$ may be different from the linear form one, the teleparallel action term.

\par
Universe is assumed to be homogeneous and isotropic at large scales. However, this is not the case if local consideration are made. In such a situation, the anisotropic effects that appears should not be described with the use of the common Friedmann-Robertson-Walker (FRW) metric. The simplest of anisotropic models usually used for describing the anisotropic  effects are Bianchi type I (BI) spatially homogeneous models whose spatial sections are flat.
The advantages in using anisotropic models are that they have significant role in the description of the early stage of the universe \cite{pitrou} and also are extremely useful in obtaining more general cosmological models that the isotropic FRW models. Anisotropy is completely supported in various angle scales by the  Differential Microwave Radiometers on NASA's Cosmic Background Explorer. Speculations indicate that the entire history of the cosmic evolution down to the recombination is hidden in anisotropies and they are conceived to be indicative the matter composing and the geometry of the universe. Important theoretical argument by Misner in 1968 and the modern experimental data defend the existence of anisotropic phase of the universe, which, later, turns into an isotropic one.
\par 
Therefore, its quite reasonable to check the isotropization of any anisotropic model  able to describe a stage of the evolution history of the universe. At a simple level, the isotropy of the present stage of the universe allows the BI model to be a prime candidate in the study of the possible effects of an anisotropy in the early universe on the current observational data. 
\par
Recently, Fadragas et al \cite{fadragas} performed a detailed dynamical analysis of anisotropic scalar-field cosmologies, and in particular of the most significant Kantowski-Sachs, Bianchi I and Bianchi III case, and as conclusion, they found   a very rich behaviour,  and amongst others the universe can result in isotropized solutions with observables in agreement with observations, such as de Sitter, quintessence-like or stiff-dark energy solutions. Fontanini et al considered the cosmic parallax effect in anisotropic cosmological models described by an axisymmetric homogeneous Bianchi I metric and discuss whether any observation of cosmic parallax would distinguish between different anisotropic evolutions \cite{fontanini}. Still in the framework of anisotropic models, Campanelli et al \cite{campanelli} analysed the magnitude-redshift data of type Ia supernovae included in the Union and Union2 compilations anisotropic Bianchi I cosmological model and in the presence of dark energy fluid anisotropic equation of state and found that the amount of deviation from isotropy of the equation of state of dark energy, the skewness, which should be restricted to the interval $(-0.16;0.12)$. Also, in order to well understand the notion of ``real-time cosmology", Quercellini et al undertook the real-time measurements of the overall redshift drift and angular separation shift in distant source, which allows the observer to trace the background cosmic expansion and large scale anisotropy, respectively \cite{quercellini2,quercellini}.  \par
Following these evident features of the universe, we focus our attention in this work to the Locally Rotationally Symmetric (LRS) BI model which is a special case of BI one. Within this metric and the general field equations give rise to three independent equations. In the sequence, we present the generalisation of the anisotropic model within torsion consideration. As we mentioned above, the isotropization is an important aspect following the evolution history of the universe. This feature is extensively analysed in the framework of LRS-BI model in this work. The final and important point undertaken in this work is the reconstruction of $f(T)$ model according to the LRS-BI metric. Across all the above studies, two interesting cases are considered. A linear form of the algebraic function and a more complicated case. Our results show that for the linear case, the real evolution of our universe holds whereas, for the more complicated one, the universe presents more anisotropy at the beginning only when the terms of higher order in $T$ fade away. \par

The paper is organized as follows. In Sec. \ref{sec2} we present the preliminary definitions and equations of motion in $f(T)$ gravity. The field equations are presented according to the LRS-BI metric in Sec. \ref{sec3}. The Sec. \ref{sec4} is devoted to the generalisation of the anisotropic models with the torsion and also to the isotropization scheme. A linear case $f(T)=0$ is developed in the Sec. \ref{sec5} and our  conclusions are presented in Sec. \ref{sec6}.

\section{ Preliminary definitions and equations of motion}\label{sec2}
As mentioned above, we define $f(T)$ theory within Weitzenkock's connection where the line element is described by 
\begin{equation}\label{el}
dS^{2}=g_{\mu\nu}dx^{\mu}dx^{\nu}\; ,
\end{equation} 
$g_{\mu\nu}$ being the components of the metric. One can describe the theory in the 
spacetime or in the tangent space. This allows  to rewrite the line
element (\ref{el}) as follows 
\begin{eqnarray}
dS^{2} &=&g_{\mu\nu}dx^{\mu}dx^{\nu}=\eta_{ij}\theta^{i}\theta^{j}\label{1}\; ,\\
dx^{\mu}& =&e_{i}^{\;\;\mu}\theta^{i}\; , \; \theta^{i}=e^{i}_{\;\;\mu}dx^{\mu}\label{2}\; ,
\end{eqnarray} 
where $\eta_{ij}=diag[1,-1,-1,-1]$ and $e_{i}^{\;\;\mu}e^{i}_{\;\;\nu}=\delta^{\mu}_{\nu}$ or  $e_{i}^{\;\;\mu}e^{j}_{\;\;\mu}=\delta^{j}_{i}$. We also express the square root of the metric determinant by  $\sqrt{-g}=\det{\left[e^{i}_{\;\;\mu}\right]}=e$ and the matrix $e^{a}_{\;\;\mu}$ are called tetrads and play the crucial role in the dynamic fields of the theory.
\par
By using theses fields, we define the Weitzenbock's connection as
\begin{eqnarray}
\Gamma^{\alpha}_{\mu\nu}=e_{i}^{\;\;\alpha}\partial_{\nu}e^{i}_{\;\;\mu}=-e^{i}_{\;\;\mu}\partial_{\nu}e_{i}^{\;\;\alpha}\label{co}\; .
\end{eqnarray}
The geometrical objects of the spacetime are then  constructed from this connection. The components of the tensor torsion are defined by the antisymmetric part of this connection
\begin{eqnarray}
T^{\alpha}_{\;\;\mu\nu}&=&\Gamma^{\alpha}_{\nu\mu}-\Gamma^{\alpha}_{\mu\nu}=e_{i}^{\;\;\alpha}\left(\partial_{\mu} e^{i}_{\;\;\nu}-\partial_{\nu} e^{i}_{\;\;\mu}\right)\label{tor}\;.
\end{eqnarray}
The components of the contorsion tensor are defined as 
\begin{eqnarray}
K^{\mu\nu}_{\;\;\;\;\alpha}&=&-\frac{1}{2}\left(T^{\mu\nu}_{\;\;\;\;\alpha}-T^{\nu\mu}_{\;\;\;\;\alpha}-T_{\alpha}^{\;\;\mu\nu}\right)\label{contor}\; .
\end{eqnarray}
In order to make more clear the definition of the scalar equivalent to the curvature scalar of RG, we first define a new tensor $S_{\alpha}^{\;\;\mu\nu}$, constructed from the components of the tensors torsion and contorsion as
\begin{eqnarray}
S_{\alpha}^{\;\;\mu\nu}&=&\frac{1}{2}\left( K_{\;\;\;\;\alpha}^{\mu\nu}+\delta^{\mu}_{\alpha}T^{\beta\nu}_{\;\;\;\;\beta}-\delta^{\nu}_{\alpha}T^{\beta\mu}_{\;\;\;\;\beta}\right)\label{s}\;.
\end{eqnarray}
We can now define the torsion scalar by the following contraction
\begin{eqnarray}
T=T^{\alpha}_{\;\;\mu\nu}S^{\;\;\mu\nu}_{\alpha}\label{te}\; .
\end{eqnarray}

The action of the theory is defined by generalizing  the Teleparallel theory, as 
\begin{eqnarray}\label{action}
S=\int \left[T+f(T)+\mathcal{L}_{Matter}\right]ed^4x\;.
\end{eqnarray}
Here, $f(T)$ denotes an algebraic function of the torsion scalar $T$. Making the functional variation of the action  (\ref{action}) with respect to the tetrads, we get the following equations of motion \cite{Baojiu1,daouda11,daouda21}
\begin{eqnarray}
S^{\;\;\nu\rho}_{\mu}\partial_{\rho}Tf_{TT}+\left[e^{-1}e^{i}_{\mu}\partial_{\rho}\left(ee^{\;\;\alpha}_{i}S^{\;\;\nu\rho}_{\alpha}\right)+T^{\alpha}_{\;\;\lambda\mu}S^{\;\;\nu\lambda}_{\alpha}\right]\;(1+\; f_{T})+\frac{1}{4}\delta^{\nu}_{\mu}\;(T+f)=4\pi\mathcal{T}^{\nu}_{\mu}\label{em}\; ,
\end{eqnarray}
where $\mathcal{T}^{\nu}_{\mu}$ is the energy momentum
tensor, $f_{T}=d f(T)/d T$ and $f_{TT}=d^{2} f(T)/dT^{2}$. By setting $f(T)= a_0 $, $a_0$ being a constant,  the equations of motion (\ref{em}) are exactly that of the Teleparallel theory with a cosmological constant, and this is dynamically equivalent to the GR. We see from this that the equations clearly depend on the choice made for the set of tetrads \cite{cemsinan}. 
\par
The energy momentum tensor incorporates the  interaction of the gravitational field with the matter ones and is defined as 
\begin{eqnarray}
\mathcal{T}^{\,\nu}_{\mu}=  diag\left(1,-\omega_x,-\omega_y,-\omega_z\right)\rho                         \label{tem}\; .
\end{eqnarray}
Here, $\omega_i$ ($i=x,y,z$) are the parameters
of equations of state related to the pressures $p_x$, $p_y$ and $p_z$.

\section{Field equations for LRS Bianchi type-I model} \label{sec3}
Let us first establish the equations of motion of a set of diagonal 
tetrads using the Cartesian coordinate metric, for  describing LRS Bianchi type-I model.  The LRS Bianchi type-I metric reads 
\begin{equation}
dS^2=dt^2-A^2(t)dx^2-B^2(t)\left(dy^2+dz^2\right)\,,\label{m}
\end{equation}
Let us choose the following set of diagonal tetrads related to 
the metric (\ref{m})
\begin{eqnarray}
\left[e^{a}_{\;\;\mu}\right]=diag\left[1,A,B,B\right]\;. \label{m1}
\end{eqnarray}
The determinant of the matrix (\ref{m1}) is $e=AB^2$. The components of the torsion tensor  (\ref{tor}) for the tetrads (\ref{m1}) are given by
\begin{eqnarray}
T^{1}_{\;\;01}=\frac{\dot{A}}{A}\,,\,T^{2}_{\;\;02}=\frac{\dot{B}}{B}\,,
\, ,\,T^{3}_{\;\;03}=\frac{\dot{B}}{B}
\;,\label{torsiontype3}
\end{eqnarray}
and the non null components of the corresponding contorsion tensor are 
\begin{eqnarray}
K^{01}_{\;\;\;\;1}=\frac{\dot{A}}{A}\,,
\,K^{02}_{\;\;\;\;2}=\frac{\dot{B}}{B}\,,
\,
,\,K^{03}_{\;\;\;\;3}=\frac{\dot{B}}{B}\;.\label{contorsiontype3}
\end{eqnarray} 
The components of the tensor $S_{\alpha}^{\;\;\mu\nu}$, in (\ref{s}), 
are given by
\begin{eqnarray}
S_{1}^{\;\;10}=\frac{\dot{B}}{B}\,,\,S_{2}^{\;\;20}=\frac{1}{2}
\left(\frac{\dot{A}}{A}+\frac{\dot{B}}{B}\right)\,,
\,S_{3}^{\;\;30}=\frac{1}{2}\left(\frac{\dot{A}}{A}+
\frac{\dot{B}}{B}\right)\;.\label{tensortype3}
\end{eqnarray}
By using the components (\ref{torsiontype3}) and (\ref{tensortype3}), 
the torsion scalar (\ref{te}) is given by
\begin{eqnarray}
T=-2\left(2\frac{\dot{A}\dot{B}}{AB}+\frac{\dot{B}^2}{B^2}\right)\;
\label{torsionScalar1}.
\end{eqnarray}

The equations of motion corresponding to LRS Bianchi type-I model are obtained by
\begin{eqnarray}
16 \pi \;\rho&=& (f+T) +4\;(1+ f_T)\left[\left(\frac{\dot{B}}{B}\right)^2+2\frac{\dot{A}\dot{B}}{AB}\right]\;,\label{densityks}\\
16 \pi \; p_x&=& (f+T) +4\; (1+ f_T) \left[\frac{\ddot{B}}{B}+\left(\frac{\dot{B}}{B}\right)^2+\frac{\dot{A}\dot{B}}{AB}\right]+4\frac{\dot{B}}{B}\dot{T}f_{TT}\;,\label{radialpressureks}
\end{eqnarray}
\begin{eqnarray}
16 \pi \; p_y&=& (f+T) +2 \;(1+ f_T) \left[\frac{\ddot{A}}{A}+\frac{\ddot{B}}{B}+\left(\frac{\dot{B}}{B}\right)^2+3\frac{\dot{A}\dot{B}}{AB}\right]+2\left(\frac{\dot{A}}{A}+\frac{\dot{B}}{B}\right)\dot{T}f_{TT}\;,\label{tangentialpressureks}\\
p_y&=&p_z\nonumber\,\,\,.
\end{eqnarray}

We parametrize it as follows
\begin{eqnarray}
\mathcal{T}^{\,\nu}_{\mu} &=&  diag\left(1,-\omega_x,-\omega_y,-\omega_z\right)\rho  \; ,\\
&=&  diag\left(1,-\omega,-(\omega + \delta),-(\omega + \delta) \right)\rho \label{tem}\,,
\end{eqnarray}
where $\rho$ is the energy density of the fluid; $p_x, \; \; p_y$ and
$p_z$
are the pressures and $\omega_x, \omega_y$ and
$\omega_z$ are the directional
equation of state (EoS) parameters of the fluid.
Now, parametrizing the deviation from isotropy by setting
\; $\omega_x=\omega$ and then introducing skewness parameter $\delta$ that
is the deviations from $ω$ respectively on both the $y$ and $z$
axes. Here $ω$ and $\delta$ are not necessarily constants and can
be functions of the cosmic time $t$.

The equations of motion corresponding to LRS Bianchi type-I model for an anisotropic fluid
are obtained by
\begin{eqnarray}
16 \pi \;\rho&=& (f+T) +4 \;(1+f_T) \left[\left(\frac{\dot{B}}{B}\right)^2+2\frac{\dot{A}\dot{B}}{AB}\right]\;,\label{ines0}\\
 -16 \pi \omega\rho&=& (f+T) +4 (1+f_T) \left[\frac{\ddot{B}}{B}+\left(\frac{\dot{B}}{B}\right)^2+\frac{\dot{A}\dot{B}}{AB}\right]+4\frac{\dot{B}}{B}\dot{T}f_{TT}\;,\label{ines1'} \\
-16 \pi \;(\omega+\delta)\rho&=& (f+T) +2\; (1+f_T) \left[\frac{\ddot{A}}{A}+\frac{\ddot{B}}{B}+\left(\frac{\dot{B}}{B}\right)^2+3\frac{\dot{A}\dot{B}}{AB}\right]+2\left(\frac{\dot{A}}{A}+\frac{\dot{B}}{B}\right)\dot{T}f_{TT}\label{ines2'}\,.
\end{eqnarray}

The conservation of the energy-momentum tensor of
the fluid parametrized in (\ref{tem}), i.e  $T_{;\nu}^{\mu \nu}=0 $, leads to
the following equation:
\begin{eqnarray}
\dot{\rho} + \Bigg[ \frac{\dot{A}}{A}(1+\omega)  + 2 \frac{\dot{B}}{B}(1+ (\omega+\delta)) \Bigg]\;\rho=0\,.\label{cons} 
\end{eqnarray}

\section{Generalized anisotropic models with Torsion and Isotropization}\label{sec4}
Before determining the parameters $\delta,\rho$ and $\omega$ 
whose  involved in the isotropization, we will determine the Hubble parameters
 in the direction of  $[x,y,z] $
\begin{eqnarray}
 H_x = \frac{\dot{A}}{A}\,,\;\;\, H_y = \frac{\dot{B}}{B} \,,\,\;\;
 H_z = \frac{\dot{B}}{B}           \label{ines1}\,.
\end{eqnarray}
The main Hubble parameter is given by 
\begin{eqnarray}
H = \frac{1}{3} \frac{\dot{V}}{V}
  &=&\frac{1}{3}\left( \frac{\dot{A}}{A} +  \frac{\dot{B}}{B} + \frac{\dot{B}}{B}\right) \cr
 &=&  \frac{1}{3} \left( \frac{\dot{A}}{A} + 2 \frac{\dot{B}}{B} \right) \label{ines2}\,,
\end{eqnarray}
with
\begin{eqnarray}
  V= A B^2  \label{tous1}\,,
\end{eqnarray} 
being the spatial volume of the universe. The rate of expansion is evaluated by anisotropy parameter given as
\begin{eqnarray}
 \Delta= \frac{1}{3} \sum^{3}_{i=1}\left( \frac{H_i - H}{H} \right)^2\,, \label{ines3}
\end{eqnarray}
where $i =(\;x \;y \;z\;)$.
\par 
By introducing    (\ref{ines2}) and (\ref{ines2}) into \ref{ines3}, the anisotropy parameter can be written as following 
\begin{eqnarray}
 \Delta= \frac{2}{9 H^2} \left( H_x - H_z \right)^2\,. \label{ines4} 
\end{eqnarray}

The scalar expansion and the shear can  be defined as 
\begin{eqnarray}
&&\theta(t)= \frac{\dot{A}}{A} + 2 \frac{\dot{B}}{B}\label{theta} \,,\\
&&\sigma (t)=\frac{1}{\sqrt{3}}\left(\frac{\dot{A}}{A} -\frac{\dot{B}}{B}\right)\label{sigma}\,.
\end{eqnarray}  

We now we have three linearly independent equations (\ref{ines0})-(\ref{cons}) and five unknown functions $ A,B,\omega,\rho$ and $\delta$. Beside to this, we make use of  an additional condition which allows us to choose the  volumetric law expansion for solving the system completely. The following two different volumetric expansion laws have been used 
laws:
\begin{eqnarray}
V= c_1 \; e^{3H_0t}\,,  \label{tous}
\end{eqnarray}
where $c_1$ and $H_0$ are two positive constants.
Because of the expansion of the universe, one can calculate the deceleration parameter as \begin{equation}
q=\left(\frac{d}{dt}H^{-1}(t)\right)-1\,.
\end{equation}

Using equations (\ref{ines0}), (\ref{ines2}), (\ref{ines4}) and the identity 
\begin{eqnarray}
H_y^2+2H_xH_y=3H^2\left[1-\frac{
\Delta(t)}{2}\right]\,,
\end{eqnarray}
the energy density can be re-written as
\begin{eqnarray}
&&3H^2=8\pi \rho_{eff}=8\pi\left(\rho+\rho_{T}+\rho_{anis}\right)\label{densE}\,,\\
&&\rho_{T}=-\frac{1}{16\pi}\left[f(T)+12H^2f_T(T)\right]
\label{densT}\,,\\
&&\rho_{anis}=\frac{3}{16\pi}H^2\Delta(t)\left[1+2f_T(T)\right]\,,\label{densA}
\end{eqnarray}
where the subscript $``anis"$ expresses the anisotropic component.
By considering the particular case $f(T)=0$, one regains the anisotropic LRS-BI model in GR. When one considers the particular case $\Delta(t)=0$ (implying $A(t)\rightarrow B(t)$ in (\ref{ines4})), $A(t)=B(t)=a(t)$, and $f(T)$ is re-obtained in the framework of flat FRW metric. If all the densities are different from zero, one gets a theory that takes into account the anisotropic contribution and the contribution of the terms non-linear in torsion scalar.
\par
In the next section, we will deal with the particular case $f(T)=0$, in order to re-obtain  
the anisotropic dark energy results, as an analogy of the GR.
\par 
Now, let us define the necessary condition for an anisotropic model of the universe presenting isotropization. First, let us explain the term ``isotropization". The isotropization is the process in which a possible anisotropic universe can adequate with the current observational data, being homogeneous and isotropic; such an universe is said becoming isotropic by the isotropization process. This process is really realised in the inflationary phase and modifies any geometry into an isotropic one and any fluid into an homogeneous one.
\par 
The criterion for having the possibility of such a model are known for several years, through the work of Collins and Hawking \cite{collins}, where they established the following conditions: when  taking the limit $t\rightarrow +\infty$, one has the function
$\delta(t)$, in (\ref{tem}), $\Delta(t)$, in (\ref{ines4}) and  $M(t)=\sigma(t)/\theta(t)$\footnote{The functions $\sigma(t)$ and $\theta(t)$ are defined in (\ref{sigma}) and  (\ref{theta}), respectively.} tends to zero. This is clear when one makes $\delta(t)\rightarrow 0$ in the definition of energy momentum tensor (\ref{tem}), obtaining a perfect fluid, and when $\Delta(t)\rightarrow 0$ in (\ref{ines4}), leading to 
$B(t)\rightarrow A(t)$, which corresponds to the flat FRW metric in (\ref{m}).

\section{The Linear Case, $f(T)=0$}\label{sec5}

For this work, we have used the model with the exponential expansion given by the relation (\ref{tous}).  
By using equations (\ref{tous1}) and (\ref{tous}), we can obtain:
\begin{eqnarray}
A (t)= \frac{c_1 \; e^{3H_0\; t}}{B^2} \label{i11}\,.
\end{eqnarray}
By making use of  (\ref{ines1'}), \; (\ref{ines2'}),\;(\ref{i11})  after solving (\ref{ines0})-(\ref{cons}), we obtain:
\begin{eqnarray}
B(t)= e^{H_0t}\left( \frac{\kappa}{H_0}  + 3\lambda \;e^{-3H_0t}\right)^{1/3} \label{i11'}\,,
\end{eqnarray}
lead to 
\begin{eqnarray}
A(t)=c_1\;e^{H_0t}\left( \frac{\kappa}{H_0}  + 3 \lambda e^{-3H_0t}\right)^{-2/3}\,. \label{i11''}
\end{eqnarray}
By introducing (\ref{i11'}) and (\ref{i11''}) into (\ref{ines2}) and (\ref{ines4}), we obtain the mean Hubble parameter 
\begin{eqnarray}
H(t)= H_0\label{H0}\,,
\end{eqnarray}
and the anisotropy parameter of the expansion as: 
\begin{eqnarray}
\Delta (t)= \frac{18\;H_0^2 \; \lambda^2 }{(\kappa\;e^{3H_0t}+ 3\; H_0\; \lambda)^2} \label{sa0}\,.
\end{eqnarray}
By using equation (\ref{ines0}), we can obtain the energy density of the fluid as:
\begin{eqnarray}
\rho_1(t)= \frac{3 \kappa\; H_0^2 \;e^{3H_0t}(\kappa\; e^{3H_0t}+6 H_0 \lambda)}{8\pi(\kappa\; e^{3H_0t}+3 H_0 \lambda)^2} \label{sa1}\,.
\end{eqnarray}
The deviated part of the anisotropic EoS parameter is obtained by using equations  (\ref{ines1}), (\ref{sa1}) as following:
\begin{eqnarray}
\omega_1(t)= - \frac{3 \kappa\; H_0^2 \;e^{3H_0t}(\kappa\; e^{3H_0t}+6 H_0 \lambda)}{8\pi(\kappa\; e^{3H_0t}+3 H_0 \lambda)^2 \; \rho_1(t)} \label{sa2}\,.
\end{eqnarray}
In particular, one has a specific value  $\omega_1(t)=-1$ in our linear case. We can determine the deviation-free part of anisotropic EoS parameter by making use (\ref{ines2}), 
(\ref{sa1}) and(\ref{sa2}) as following:
\begin{eqnarray}
\delta_1(t)= - \frac{27 H_0^2 \; \lambda^2 \;e^{-3H_0t} }{\kappa(\kappa e^{3H_0t} + 6H_0\lambda)} \label{sa3}\,.
\end{eqnarray}

One can calculate the functions $\theta(t)$ and $\sigma(t)$, with  (\ref{theta}) and  (\ref{sigma}), yielding 
\begin{eqnarray}
\theta(t)=3H_0\,,\,\sigma(t)=\frac{3\sqrt{3}H_0^2 \lambda}{3H_0\lambda+\kappa e^{3H_0 t}}\label{theta-1}\,.
\end{eqnarray}
\par 
This result can be interpreted as an universe in an accelerated expansion $q=-1$, with a constant mean Hubble parameter and a volume involving with the cosmic time $t$, as shown in the Figure $1$. Due to the fact that $H(t)=H_0$, on has an universe of type de Sitter, where the effective density in (\ref{densE}) is constant, mimicking the cosmological constant.  
\par
The scale factors in the directions $x$ and $y$ (also $z$) are increasing, as represented in the Figure $2$. At the beginning, the universe increases more rapidly in the direction $y$, and later  being equal to the others (isotropization). \par
\begin{figure}[htbp]
\begin{center}
\includegraphics[width=6cm, height=5cm]{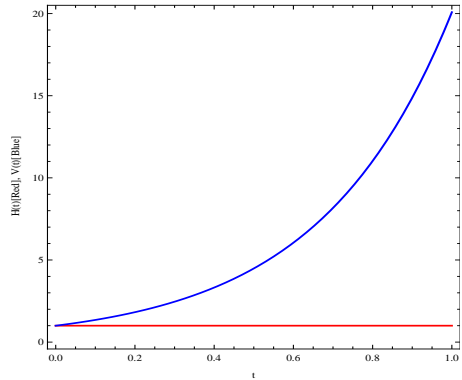}\label{V}
\end{center}
\caption{\small{The graphs illustrating the evolution of the average Hubble parameter $H(t)$ and the volume $V(t)$ of the universe. The functions are plotted for $c_1=1$ and $H_0=1$. }}
\end{figure}
\begin{figure}[htbp]
\begin{center}
\includegraphics[width=6cm, height=5cm]{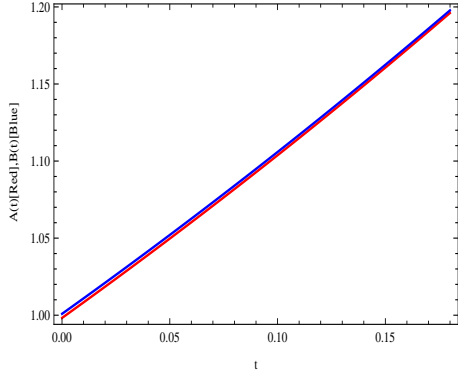}\label{A}
\end{center}
\caption{\small{The graphs illustrating the evolution of the scale factors $A(t)$ and $B(t)$. The functions are plotted for $c_1=1$,\;\; $H_0=1$, \;\;$ T_0=1$, \;\;$ T_2=1$, \;\;$ \lambda=10^{-3}$ \;and $\;\; \kappa=0.9996711436381983$. }}
\end{figure}
This is an isotropization model because the functions  $\delta_1(t)$, in (\ref{sa3}), and  $\Delta(t)$, in (\ref{sa0}), tend to zero as the cosmic time goes toward infinity  ($p_x=p_y=p_z=p=\omega \rho$ and $A(t)=B(t)$). They are the anisotropic contributions for this model. The functions $\theta(t),\sigma(t)$ and  $\Delta(t)$ are described in the Figure $3$. One can see in these expressions that the isotropization is obtained at the limit $\lambda\rightarrow 0$. We clearly see that it is the parameter $\lambda$ which governs the isotropization of the universe in this case. If  $\lambda$ goes to infinity in (\ref{sa0}), one gets the maximal anisotropy, so that  $\Delta_{Max}=2$. Hence, $\Delta\leq 2$ in our case. 
\begin{figure}[h]
\centering
\begin{tabular}{rl}
\includegraphics[height=3cm,width=5cm]{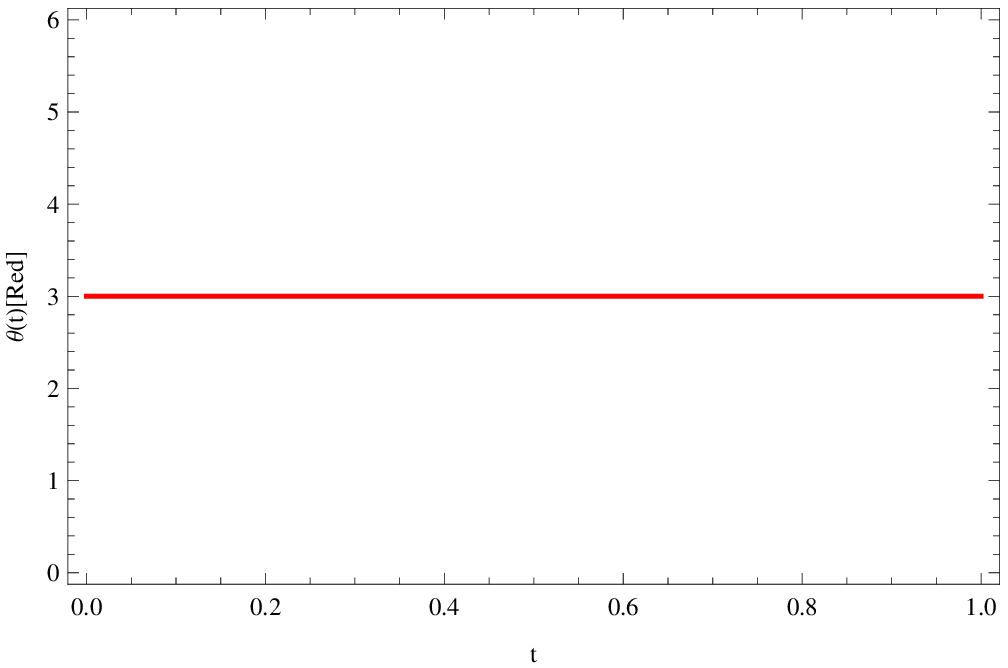}&
\includegraphics[height=3cm,width=5cm]{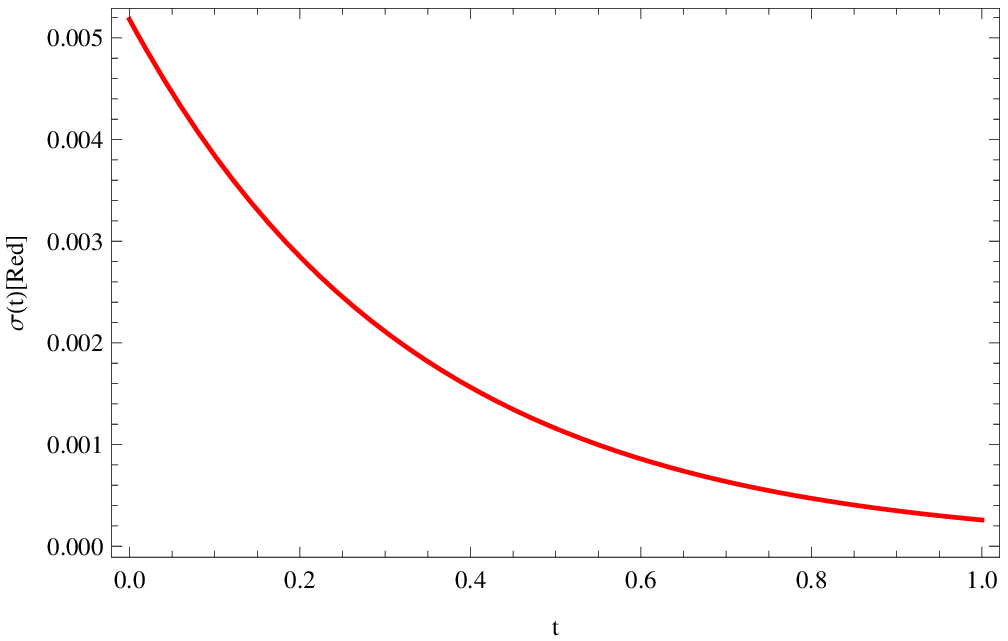}\\
\includegraphics[height=3cm,width=5cm]{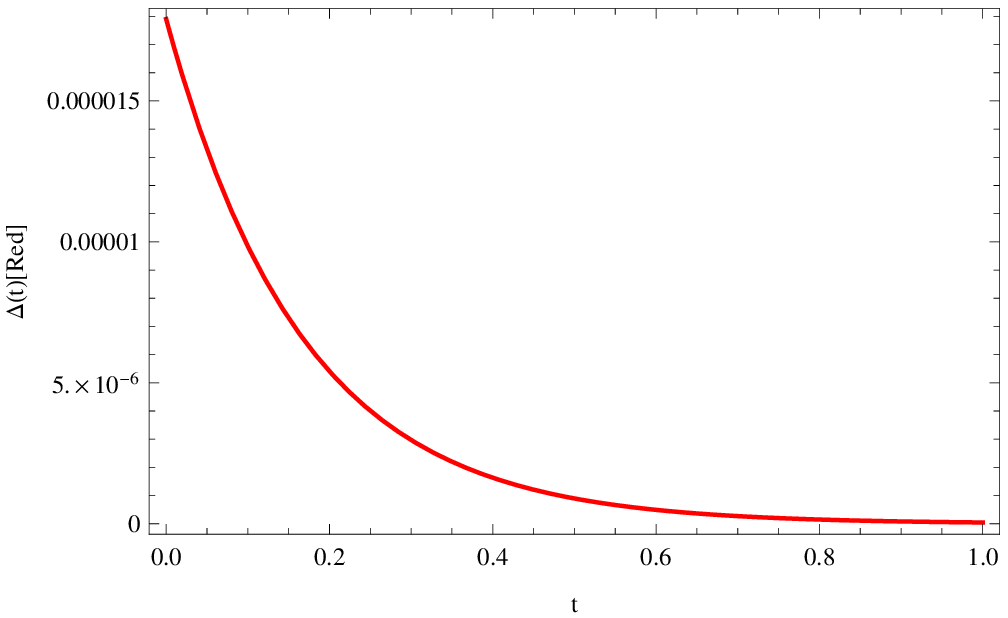}
\end{tabular}
\caption{\scriptsize{The graphs illustrating the  evolution of the scalar expansion $\theta(t)$, the shear scalar $\sigma(t)$ and the anisotropy parameter $\Delta(t)$. The functions are plotted for $c_1=1$,\;\; $H_0=1$, \;\;$\lambda=10^{-3}$ and \;\; $\kappa=0.9996711436381983$.}} \label{fig2}
\end{figure}

\par
One can define a function which gives the anisotropic measure by
\begin{eqnarray}
M(t)=\frac{\sigma(t)}{\theta(t)}=\frac{\sqrt{3}H_0\lambda}{3H_0\lambda+\kappa e^{3H_0 t}}\label{M}\,.
\end{eqnarray}
The curve that shows the behaviour of the function  $M(t)$ is described by the  Figure $4$, so that, the anisotropy fades away when the time goes to infinity.
\begin{figure}
\begin{center}
\includegraphics[width=6cm, height=5cm]{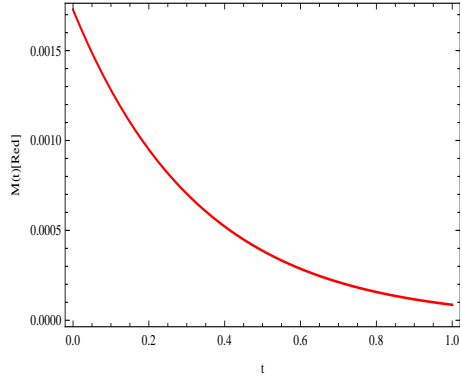}\label{fig-M}
\end{center}
\caption{\small{The graphs illustrating the  evolution of the function  $M(t)$. The function is plotted for $c_1=1$,\;\; $H_0=1$, \;\;$ T_0=1$, \;\; $T_2=1\;$, \;\;$ \lambda=10^{-3}$ and  \;\; $\kappa=0.9996711436381983$.}}
\end{figure}
\par 
In the next section we will perform the reconstruction of the function $f(T)$ for a specific case of an anisotropic geometry. The analysis of the matter content of this present section also will be done in the next section for a whole comparison.

\section{Reconstruction for LRS Bianchi type-I  model in $f(T)$ Gravity}\label{sec6}
Here, we will consider that the anisotropic geometry is given by the same functions  $A(t)$ and $B(t)$ in (\ref{i11'}) and (\ref{i11''}). Hence, we will get modifications in the matter content of the universe (the anisotropy of the fluid and the new contribution of the non-linear terms of $T$).  One has the same functions for the volume  (\ref{tous}), the mean Hubble parameter (\ref{H0}), the scalar expansion and the shear in  (\ref{theta-1}). Therefore, the model is clearly an isotropization model for the universe.
\par
For finding a good model for this theory, we will use the so-called reconstruction method, because the algebraic function $f(T)$ is reconstructed according to the geometry and the matter content of the universe. 
\par 
By introducing  (\ref{i11'}) and (\ref{i11''})
into (\ref{ines0}), we obtain the energy density of the fluid for the non-linear case 
\begin{eqnarray}
\rho_2(t)= \frac{1}{16\; \pi} \Big[6H_0^2 + f + 12 H_0^2 \;f_T - \frac{54H_0^4\; \lambda^2 \; (1+2\;f_T)}{(e^{3H_0t} \kappa + 3H_0 \lambda)^2} \Big]\label{s1}\,.
\end{eqnarray}
After solving (\ref{ines1}) and (\ref{ines2}), we obtain the deviation-free part of anisotropic EoS parameter for the non-linear case:
\begin{eqnarray}
&&\omega_2(t)= \frac{1}{16\; \pi \; \rho_2(t)} \Big\{-6H_0^2 + \frac{54H_0^4\; \lambda^2 }{(e^{3H_0t} \kappa + 3H_0 \lambda)^2}-f-12H_0^2 \left[1-\frac{9H_0^2\; \lambda^2 }{(e^{3H_0t} \kappa + 3H_0 \lambda)^2}  \right]\;f_T \nonumber\\
&&+ 
\frac{1296H_0^6 \; \kappa^2 \;\lambda^2 \; f_TT \;e^{6H_0 t}}{(e^{3H_0t} \kappa + 3H_0 \lambda)^4} \Big\}
 \label{s2}\,,
\end{eqnarray}
and the skewness parameter for the non-linear case  as following
\begin{eqnarray}
 \delta_2(t) &=& \frac{1}{16\; \pi \; \rho_2(t)} \Big[-6H_0^2-f-12H_0^2\; f_T- \frac{5832 H_0^8 \;\lambda^4 \; f_{TT}}{(e^{3H_0t} \kappa + 3H_0 \lambda)^4 } -
 \frac{1944 H_0^7 \;\lambda^3 \; f_{TT}}{(e^{3H_0t} \kappa + 3H_0 \lambda)^3 }  \cr
&& - \frac{ 54 H_0^4 \; \lambda^2 \;(2+ f_T -24 H_0^2\; f_{TT})}{(e^{3H_0t} \kappa + 3H_0 \lambda)^2} - 16\; \pi\; \rho_2(t)\omega_2(t) \Big] \label{s3}\,.
\end{eqnarray}
By replacing (\ref{i11'})
and (\ref{i11''}) into (\ref{torsionScalar1}), we obtain the following relation:
\begin{eqnarray}
 t(T)= \frac{1}{H_0}\; \log \left[\left(\frac{3\lambda H_0}{\kappa}\right)^{1/3}\; \left( \frac{\sqrt{6}H_0}{\sqrt{6H_0^2+ T}} -1\right)^{1/3} \right]       \label{s6}\,,
\end{eqnarray}
which leads to the  equation of conservation for energy-momentum tensor 
\begin{eqnarray}
- \frac{52488 H_0^{10} \;\lambda^5 \; [ (6H_0^2+ T) - \sqrt{6}H_0\sqrt{(6H_0^2+ T)}][f'(T)-f''(T)]}{\;(6H_0^2+ T)^2} =0\,,
\end{eqnarray}
where the solution is given by:
\begin{eqnarray}
 f(T)= T_0 + T_1 \; e^{T} \label{m}\,,
\end{eqnarray}
where $T_0$ and $T_1$ are integration constants. The equations  (\ref{s1}) and (\ref{s2}) become 
\begin{eqnarray}
\rho_2(t)&=& \frac{1}{16\pi}  \Big[T_0 + T_1 \; \exp\left[-\frac{6 H_0^2 \; \kappa e^{6H_0t}\; (e^{3H_0t} \kappa + 6H_0 \lambda)}{(e^{3H_0t} \kappa + 3H_0 \lambda)^2}\right]\nonumber\\&& + 
H_0^2 \Big(6+ 12 T_1\exp\left[-\frac{6 H_0^2 \; \kappa e^{6H_0t}\; (e^{3H_0t} \kappa + 6H_0 \lambda)}{(e^{3H_0t} \kappa + 3H_0 \lambda)^2}\right]\nonumber\\
&&+ \frac{54H_0^4\lambda^2 \left(-1-2  T_1 \exp\left[-\frac{6 H_0^2 \; \kappa e^{6H_0t}\; (e^{3H_0t} \kappa + 6H_0 \lambda)}{(e^{3H_0t} \kappa + 3H_0 \lambda)^2}\right]\right) }{(e^{3H_0t} \kappa + 3H_0 \lambda)^2}\Big)\Big],
         \label{s1'}
\end{eqnarray}
\begin{eqnarray}
 \omega_2(t)=- \frac{X}{\alpha} \label{s1''}\,,
\end{eqnarray}
where
\begin{eqnarray}
&&X = \Big[6H_0^2 + T_0 +   T_1\exp\left[-\frac{6 H_0^2 \; \kappa e^{6H_0t}\; (e^{3H_0t} \kappa + 6H_0 \lambda)}{(e^{3H_0t} \kappa + 3H_0 \lambda)^2}\right]
- \cr && \frac{ 1296 T_1 H_0^6 \kappa^2 \lambda^2 \exp[- \frac{6H_0 \Big(e^{6H_0t}(H_0-t) \kappa^2 + 6 H_0\kappa \lambda e^{3H_0t} (H_0-t) -9H_0^2t \lambda^2 \Big)}{(e^{3H_0t} \kappa + 3H_0 \lambda)^2}] }{(e^{3H_0t} \kappa + 3H_0 \lambda)^4}  - \frac{54 H_0^4 \lambda^2 }{(e^{3H_0t} \kappa + 3H_0 \lambda)^2} + \cr
&& 12H_0^2\; \kappa \;T_1 \;(e^{3H_0t} \kappa + 6H_0\; \lambda) \exp[- \frac{6H_0 \Big(e^{6H_0t}(H_0-t) \kappa^2 + 6 H_0\kappa \lambda e^{3H_0t} (H_0-t) -9H_0^2t \lambda^2 \Big)}{(e^{3H_0t} \kappa + 3H_0 \lambda)^2}]                                                         \Big],
\end{eqnarray}
\begin{eqnarray} 
&&\alpha = T_0+  T_1 \exp\left[-\frac{6 H_0^2 \; \kappa e^{6H_0t}\; (e^{3H_0t} \kappa + 6H_0 \lambda)}{(e^{3H_0t} \kappa + 3H_0 \lambda)^2}\right]\cr
&& + H_0^2 \Big(6+12  T_1\exp\left[-\frac{6 H_0^2 \; \kappa e^{6H_0t}\; (e^{3H_0t} \kappa + 6H_0 \lambda)}{(e^{3H_0t} \kappa + 3H_0 \lambda)^2}\right]\cr
&&  + \frac{54 H_0^4 \lambda^2}{(e^{3H_0t} \kappa + 3H_0 \lambda)^2}\Big(-1-2  T_1\exp\left[-\frac{6 H_0^2 \; \kappa e^{6H_0t}\; (e^{3H_0t} \kappa + 6H_0 \lambda)}{(e^{3H_0t} \kappa + 3H_0 \lambda)^2}\right]\Big).
\end{eqnarray} 
The expression for the skewness parameter (\ref{s3}) is given by 
\begin{eqnarray}
 \delta_2(t)=- \frac{Z}{Q} \label{s1'''},
\end{eqnarray}
where
\begin{eqnarray}
 &&Z  = 162 H_0^4 \lambda^2 \;\Big(\kappa^2\;  \exp\left[\frac{6H_0 ( \kappa^2(H_0+t) e^{6H_0t}\;+ 6H_0 \;(H_0+t)e^{3H_0t} \lambda \kappa+ 9\;tH_0^2 \;\lambda^2)}{(e^{3H_0t}\; \kappa + 3H_0 \lambda)^2}\right] \cr
 &&+ T_1 \kappa^2  e^{6H_0t}+ 6 H_0 \;\kappa \; \lambda \;
  \exp\left[\frac{3H_0 ( \kappa^2(2H_0+t) e^{6H_0t}\;+ 6H_0 \;(2H_0+t)e^{3H_0t} \lambda \kappa+ 9\;tH_0^2 \;\lambda^2)}{(e^{3H_0t}\; \kappa + 3H_0 \lambda)^2}\right]\cr
  && - 6H_0\kappa\lambda T_1  e^{3H_0t}(6H_0^2-1)+ 9H_0^2 \lambda^2 \exp\left[-\frac{6 H_0^2 \; \kappa e^{6H_0t}\; (e^{3H_0t} \kappa + 6H_0 \lambda)}{(e^{3H_0t} \kappa + 3H_0 \lambda)^2}\right]+9H_0^2  T_1\lambda^2 \Big),
\end{eqnarray}
 and 
\begin{eqnarray}
 &&Q = (e^{3H_0t} \kappa + 3H_0 \lambda)^2
 \Big(\kappa^2 (6H_0^2+T_0)   \exp\left[\frac{6H_0 ( \kappa^2(H_0+t) e^{6H_0t}+ 6H_0(H_0+t)e^{3H_0t} \lambda \kappa+ 9tH_0^2 \lambda^2)}{(e^{3H_0t}\kappa + 3H_0 \lambda)^2}\right] \cr
 && + T_1 \kappa^2 e^{6H_0t}(1+ 12H_0^2)+  6H_0\kappa\lambda (6H_0^2+T_0)\exp\left[\frac{3H_0 ( \kappa^2(2H_0+t) e^{6H_0t}+ 6H_0 (2H_0+t)e^{3H_0t} \lambda \kappa+ 9tH_0^2\lambda^2)}{(e^{3H_0t} \kappa + 3H_0 \lambda)^2}\right] \cr
&&+ 6H_0 \kappa \lambda T_1 e^{3H_0t}(1+12H_0^2)+ 9H_0^2\lambda^2 T_0  \exp\left[\frac{6 H_0^2 \kappa e^{3H_0t}(e^{3H_0t} \kappa + 6H_0 \lambda)}{(e^{3H_0t} \kappa + 3H_0 \lambda)^2}\right]+ 9H_0^2 T_1 \lambda^2 \Big).  
\end{eqnarray}
\par 
With the expressions of the energy densities  (\ref{sa1}) and (\ref{s1'}), one represents this function in the Figure $5$. Now, one can interpret the densities contributions.  The density  $\rho_1(t)$ contributes lower than the energy density  $\rho_2(t)$, for the anisotropic expansion of the universe. Indeed, one calculates the diverse contributions of the effective energy densities in (\ref{densE}). 
\begin{figure}[h]
\centering
\begin{tabular}{rl}
\includegraphics[height=3cm,width=5cm]{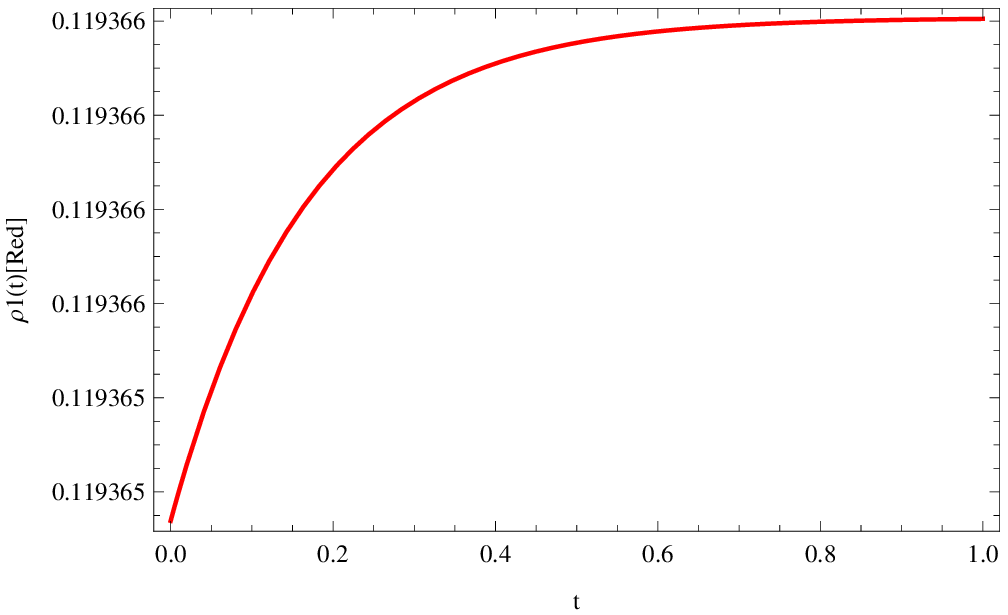}&
\includegraphics[height=3cm,width=5cm]{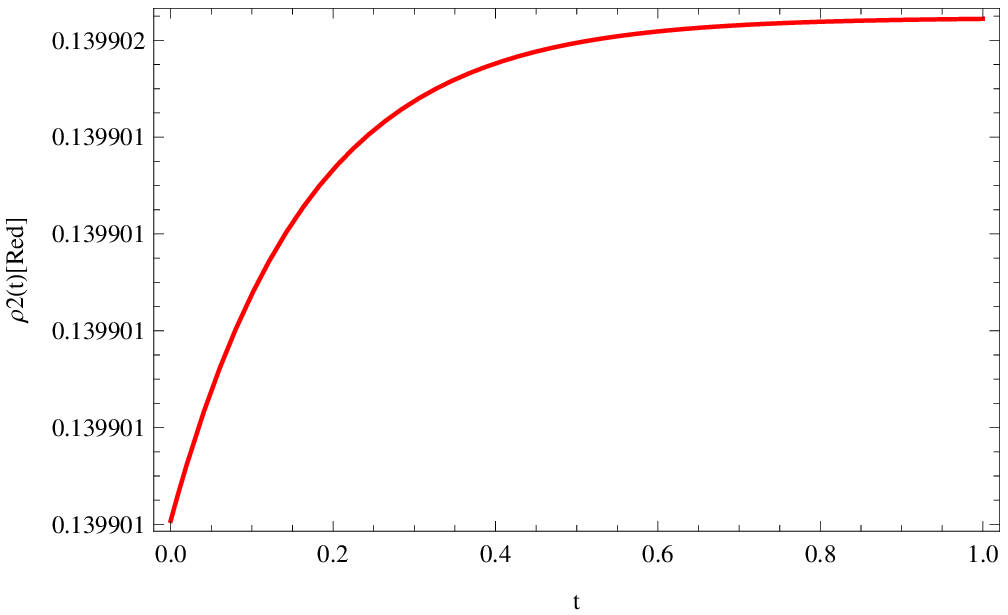}
\end{tabular}
\caption{\scriptsize{The graphs illustrating the  evolution of the energy density for the linear and non-linear cases ($\rho_1(t)$ and $\rho_2(t)$). The functions are plotted for $c_1=1$,\;\; $H_0=1$ \;\; $T_0=$1 \;\; $T_2=1$, \;\;$\lambda=10^{-3} \;\;$ and $\kappa=0.9996711436381983$.}} \label{fig2}
\end{figure}
With  (\ref{H0}), (\ref{sa0}), (\ref{m}), (\ref{densT}) and (\ref{densA}), we have the density due to the torsion scalar terms and the density due to the anisotropy. 
\begin{eqnarray}
&&\rho_{T(2)}(t)= -\frac{1}{16\pi}\exp\left[-\frac{6H_0^2\kappa e^{3H_0 t}(\kappa e^{3H_0 t}+6H_0\lambda)}{(\kappa e^{3H_0 t}+3H_0\lambda)^2}\right]\label{densT-2},\\
&&\rho_{A(2)}(t)= \frac{27H_0^4\lambda^2}{8\pi}\frac{\left(1+2T_1\exp[-\frac{6H_0^2\kappa e^{3H_0 t}(\kappa e^{3H_0 t}+6H_0 \lambda)}{(\kappa e^{3H_0 t}+3H_0 \lambda)^2}]\right)}{(\kappa e^{3H_0 t}+3H_0 \lambda)^2}\label{densA-2}.
\end{eqnarray}
The densities of the linear case are given by 
\begin{eqnarray}
&&\rho_{T(1)}(t)= 0\label{densT-1},\\
&&\rho_{A(1)}(t)=\frac{27H_0^4\lambda^2}{8\pi(\kappa e^{3H_0 t}+3H_0 \lambda)^2} \label{densA-1}.
\end{eqnarray}
We represent the curves of the energy densities  $\rho_{A(1)}(t)$ and  $\rho_{A(2)}(t)$ at the left hand side in the Figures $6$.  We recall that at the left hand side there are two overlapping curves despite of seeming have one. In order to clearly show that there are effectively two different  curves, within the same parameters and plotting in the range $t\in (0,0.02)$, the curve at the right hand side, it appears clearly that the two curves are distinct. One sees directly that the most important contribution is that of the model $2$, i.e, the non-linear in $T$.  This is due to the negative contribution of the energy density  $\rho_{T(2)}$, which is different from zero for the non-linear case  (see Figure $7$). So, the universe is more anisotropic at the beginning if the term of higher order on the torsion scalar are not considered. This is the main and interesting result obtained here, which is closest to the observational data.  
\par
Moreover, the energy densities combine, giving rise to a constant effective energy density, for getting a universe of type de Sitter. The two models do not interact.


\begin{figure}[h]
\centering
\begin{tabular}{rl}
\includegraphics[height=3cm,width=5cm]{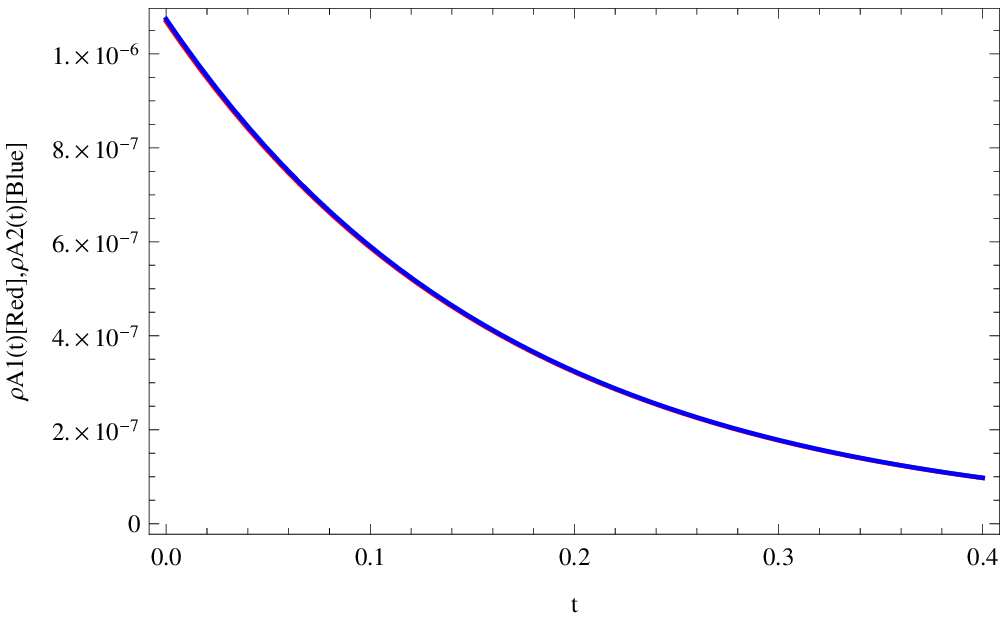}&
\includegraphics[height=3cm,width=5cm]{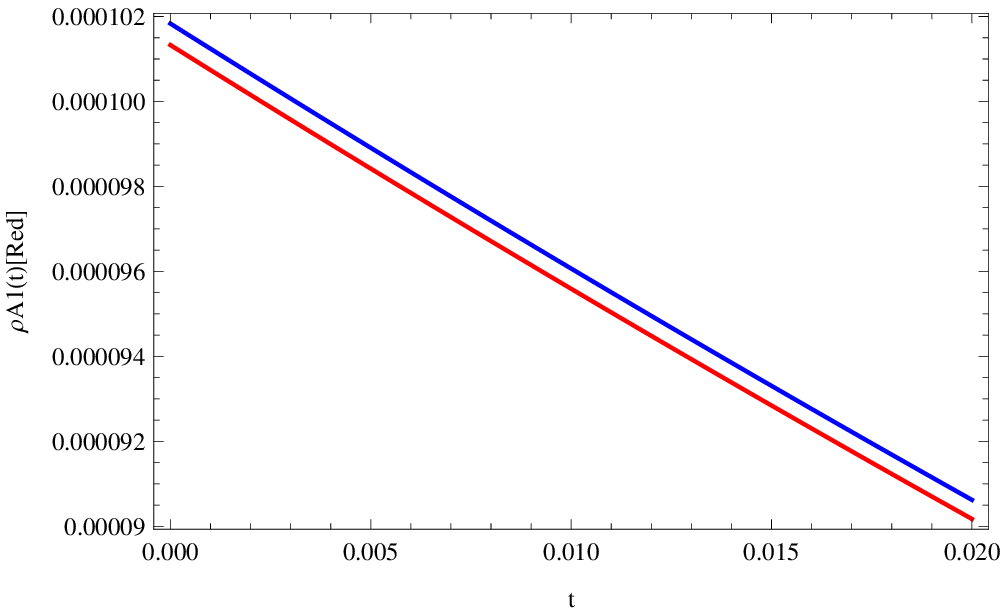}
\end{tabular}
\caption{\scriptsize{The graphs at the left hand side illustrates the  evolution of the energy densities in (\ref{densA-2}) and (\ref{densA-1}), for the linear and non-linear cases in the inflationary phase, showing the decaying of the curves toward zero; while, the one at  the right hand side exhibits the different between the curves. The functions are plotted for $c_1=1$,\;\; $H_0=1$, \;\; $T_0=1$, \;\; $T_2=1$, \;\;$\lambda=10^{-3}$ and $\kappa=0.9996711436381983$.}}
\label{fig2}
\end{figure}
\par

\begin{figure}[h]
\centering
\begin{tabular}{rl}
\includegraphics[height=3cm,width=5cm]{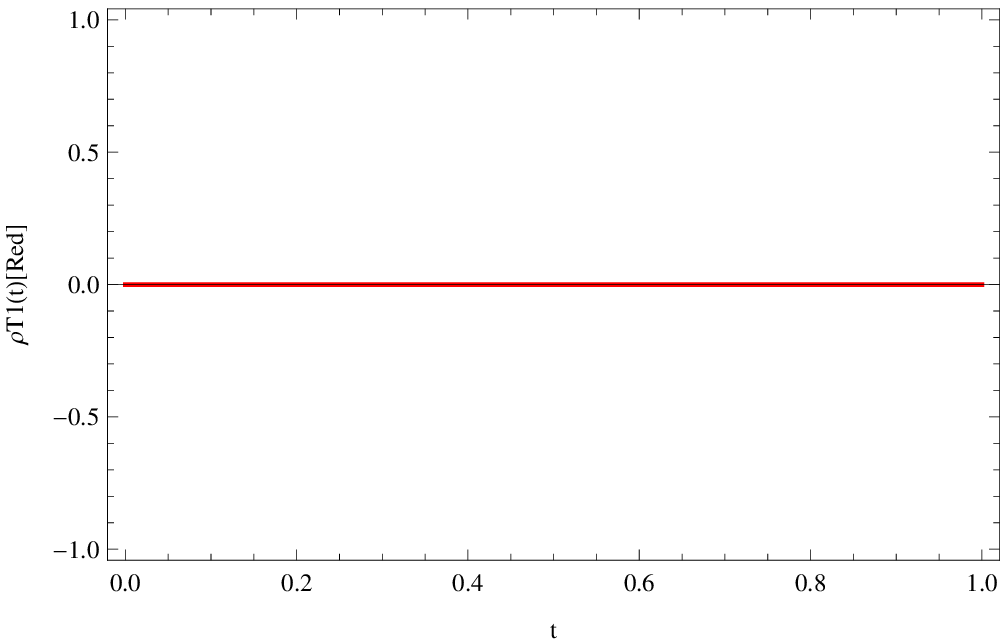}&
\includegraphics[height=3cm,width=5cm]{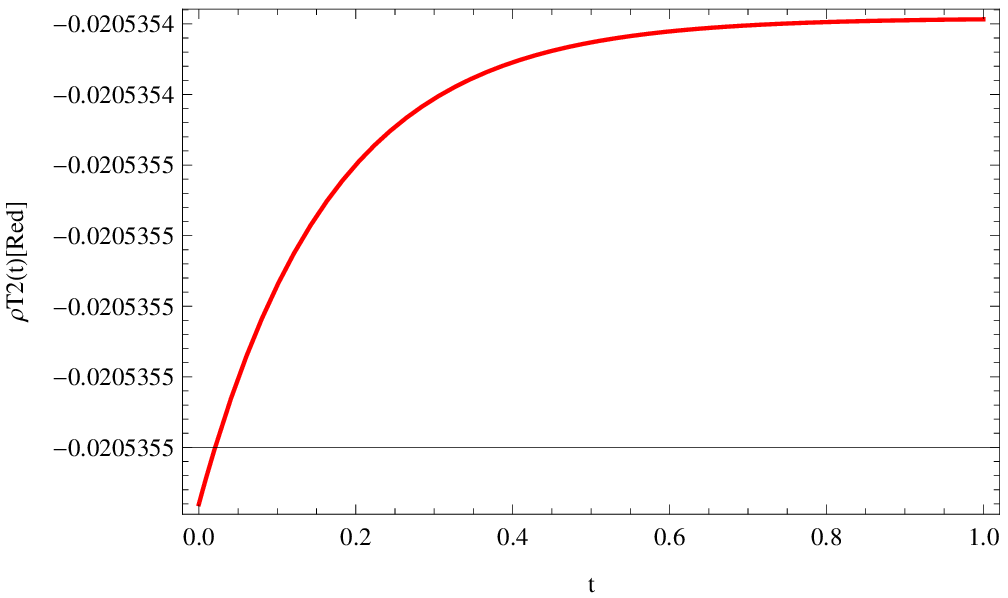}
\end{tabular}
\caption{\scriptsize{The graphs illustrating the  evolution of the energy densities in (\ref{densT-2}) and (\ref{densT-1}), for the linear and non-linear $c_1=1$,\;\; $H_0=1$, \;\; $T_0=1$, \;\; $T_2=1$, \;\;$\lambda=10^{-3}$ and $\kappa=0.9996711436381983$.}} \label{fig2}
\end{figure}
\par 
One can see very clearly this result through the figures $8$ and $9$. We see that the  skewness parameter is negative in all two cases, however, lower absolute values for the second. This comes to show us that, due to the contribution of the non-linear terms of the function $f(T)$, the parameter $\delta_2(t)$ is more restricted, for small values within the real interval $(-0.16,0.12)$ \cite{campanelli} than the parameter $\delta_1(t)$ arising from the linear theory. So, we have the model with non-null $ f(T)$, non-linear case, which is favoured by observational data, compared to the linear case.
\par 
Also, in the same way, the parameter of EoS $\omega_1(t)$ is negative, but initially less than the case of $\omega_2(t)$.       
\begin{figure}[htbp]
\begin{center}
\includegraphics[width=6cm, height=5cm]{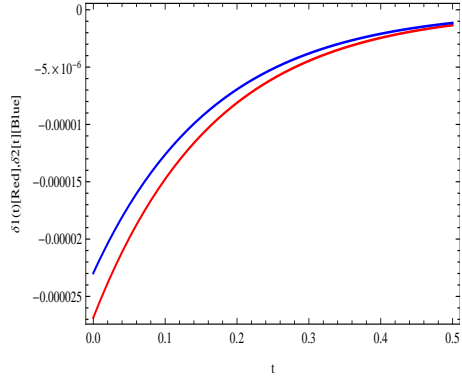}\label{delta-3}
\end{center}
\caption{The graphs illustrating the  evolution of the skewness parameter for the linear and non-linear cases ($\delta_1(t)$ and $\delta_2(t)$). The functions are plotted for $c_1=1$,\;\; $H_0=1$, \;\; $T_0=1$, \;\; $T_2=1$, \;\;$\lambda=10^{-3}$ and  \;$\kappa=0.9996711436381983$.}
\end{figure}
\begin{figure}[htbp]
\begin{center}
\includegraphics[width=6cm, height=5cm]{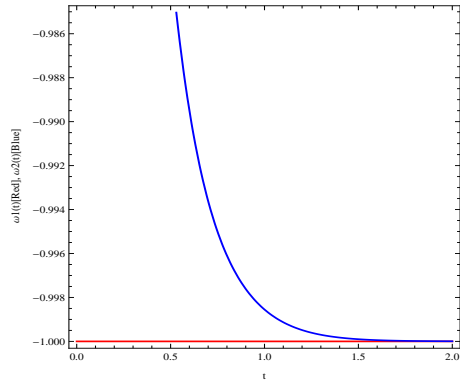}\label{omega-3}
\end{center}
\caption{The graphs illustrating the  evolution of the EoS parameters for the linear and non-linear cases ($\omega_1(t)$ and $\omega_2(t)$). The functions are plotted for $c_1=1$, \;\;$H_0=1$, \;\;$T_0=1$, \;\;$T_2=1$, \;\;$\lambda=1$\;\; and \;\;$\kappa=1$.}
\end{figure}
\par 
In order to confirm our result, let us define, as in \cite{barrow3},  the sum of the  skewness parameters
\begin{eqnarray}
\Delta_1(t)=2\delta_1(t)\,,\,\Delta_2(t)=2\delta_2(t)\,,
\end{eqnarray}  
which, from the dominant  and strong-energy conditions \cite{barrow3},  leads to the restriction 
\begin{eqnarray}
-3\leq \Delta_{1,2}\leq 3\,.\label{condDel}
\end{eqnarray}
We just have to observe the figure $8$ ( multiplying by two), for seeing that both the models obey the restriction  (\ref{condDel}), i.e, $\Delta_{1,2}\lesssim 10^{-5}$.
\par 
We can also observe the magnitude of $\sigma(t)$, normalized by the average Hubble parameter, as in \cite{koivisto}. The restriction for this value is within the same range $(-3,3)$ \cite{koivisto}. By the figure $3$, we see that $R(t)=\sqrt{3}\sigma(t)/H(t)\lesssim 10^{-2}$, which confirms the validity of the two models.

\par 
We conclude defining the  eccentricity as
\begin{eqnarray}
e_{y}^2=\left[\frac{A(t)}{B(t)}\right]^2-1\,,
\end{eqnarray}
and the contribution of the quadrupole for the anisotropies of cosmic microwave background as being \cite{koivisto}
\begin{eqnarray}
Q_2=\frac{2}{5\sqrt{3}}e^{2}_{y}\label{quadr}\,.
\end{eqnarray}
Using (\ref{i11'}), (\ref{i11''}) and  (\ref{quadr}), one gets  
\begin{eqnarray}
Q_2=\frac{2}{5\sqrt{3}}\left[\frac{c_1^2}{\left(\kappa/H_0+3\lambda e^{-3H_0t}\right)^{2}}-1\right]\label{quadr1}\,.
\end{eqnarray}
We present the graph of  $Q_2(t)$ in the figure $10$. We see that both the models obey the restriction $Q_2(t)\lesssim 10^{-6}$. Then, theses models are in agreement with the observational data, which restrain to  $Q_2(t)\lesssim 10^{-5}$. 
\begin{figure}[htbp]
\begin{center}
\includegraphics[width=6cm, height=5cm]{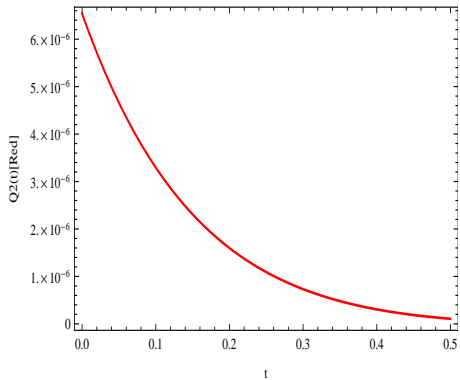}\label{Q}
\end{center}
\caption{The graphs illustrating the  evolution of the $Q_2(t)$ in (\ref{quadr1}). The functions are plotted for $c_1=1$,\;\; $H_0=1$, \;\; $T_0=1$, \;\; $T_2=1$,\;\; $\lambda=10^{-3}$ and  $\kappa=0.9996711436381983$.}
\end{figure}
\newpage
\section{Conclusion}\label{sec7}
In this work, locally rotationally symmetric Bianchi type-I model is considered in the framework of $f(T)$ theory of gravity where $T$ denotes the torsion scalar. With the corresponding metric, we wrote down the full equations of motion. We extensively performed the generalisation of anisotropic models with torsion and put a special emphasis on the isotropization process. Isotropization is an important aspect in cosmology and needs to receive particular attention, since in it evolution history, universe presents anisotropies at early time and comes becoming isotropic at present time according to the observational data. In this way, we first consider the linear case $f(T)=0$, where undertook the exponential expansion for scale factors. Then, we calculated the deviation-free part of the anisotropic EoS and the result reflects an accelerated expansion of the  universe with $q=-1$, and a mean Hubble parameter and volume involving with time. Our analysis shows that at the beginning the universe presents unequal scalar factor in the directions $x$ and $y$ (anisotropy), and as the late-time is reached, they become equal (isotropy), reflecting the isotropization process.
\par
Another interesting feature undertaken in this work is the reconstruction scheme of the algebraic function $f(T)$. We assumed the previous anisotropy expression of the scale factor and search for the action for which this anisotropy is realized. The result shows that the algebraic function is an exponential one. A whole comparison of these results with the linear case ones, as an important and interesting conclusion, is that the universe is more anisotropic at the beginning if the terms of higher order in torsion scalar are not considered. This means that the non-linear model should be favoured by observational data.
\vspace{0.2cm}

{\bf Acknowledgement}:  M. E. Rodrigues thanks a lot PPGF of the UFPA for the hospitality during the elaboration of this work and also CNPq for financial support.  I. G. SALAKO thanks  ICTP/IMSP for partial financial support.


\end{document}